\documentstyle[aps,prl,epsf]{revtex} \twocolumn
\newcommand{\DE}{{\scriptstyle\Delta}E}
\newcommand{\HH}{{\cal H}}
\newcommand{\ee}[1]{{\mbox{e}^{#1}}}

\newcommand{\Eaverage}[1]{\langle {#1} \rangle}

\newcommand{\np}{}
\begin{document}
\title{Phase Coherence in a Random One-Dimensional System of
 Interacting Fermions:
 A Density Matrix Renormalization Group Study}
 \author{Peter Schmitteckert and Ulrich Eckern}
 \address{Institut f\"ur Physik, Universit\"at Augsburg,  
          D-86135 Augsburg, Germany}
\date{\today}
\maketitle
%
\begin{abstract}
Using the density matrix renormalization group algorithm (DMRG), we study     
the model of spinless fermions with nearest-neighbor interaction
on a ring in the presence of disorder.
We determine the spatial decay of the density induced by a defect 
(Friedel oscillations),
and the phase sensitivity of the ground state energy 
$\DE= (-)^{N} (E(\phi=0) - E(\phi=\pi))$,
where $\phi = 2\pi \Phi/\Phi_0$ 
($N$ is the number of fermions,
 $\Phi$ the magnetic flux, and $\Phi_0=h/e$ the flux quantum),
for a disordered system versus the system size $M$.
The quantity $\ln{(M \DE)}$ is found to have a normal distribution to a good 
approximation.
The ``localization length'' decreases (increases) 
for a repulsive (attractive) interaction.
\end{abstract}
%
%
%
\section*{Introduction}
About five decades after the first theoretical discussion of orbital 
magnetism of free electrons on a ring \cite{Hund38}, this phenomenon
--- or ``persistent'' current -- has been observed in mesoscopic
metallic \cite{Levy90,Chandra91} and semiconducting \cite{Mailly93,Reulet95}
ring structures.
At present, the magnitude of the effect 
--- experimental results are much larger than theory predicts ---
is not well understood.
It was suggested immediately after the experiment that the constraint of
local charge neutrality, imposed by the
electron-electron interaction, could lead to a considerable enhancement
of the current \cite{Ambega90,Schmid91}, but these results, 
at least for single rings, were not conclusive \cite{Eckern92},
and were not confirmed numerically \cite{Abraham93}
(see also e.g.\ \cite{Altshul92,Kopietz93}).
\par
Considering electrons on a ring, the ground state energy, $E(\phi)$,
depends on the boundary condition, 
characterized by the phase $\phi$ 
($\phi=0$ corresponds to periodic, and $\phi=\pm\pi$ to 
antiperiodic boundary conditions).
Alternatively, the boundary condition can be interpreted as arising 
from a magnetic flux, 
$\Phi$, provided we identify $\phi$ with $2\pi\Phi/\Phi_0$, where 
$\Phi_0=h/e$ denotes the flux quantum.
Clearly, $E(\phi)$ is periodic with period $2\pi$.
The energy difference between periodic and antiperiodic boundary
conditions, $\DE$,
the persistent current, $I\sim -E'(\phi)$,
and the charge stiffness (the ``Drude weight''), $D\sim E''(\phi=0)$,
are a measure of the phase sensitivity of the system
(a comprehensive discussion is given in \cite{Giamarchi95}).
In view of the discrepancy between experiment and theory,
in particular for the metallic samples
\cite{Levy90,Chandra91}, it is important to understand further the
interplay between interaction and disorder.
Progress has been made for simple one-dimensional models \cite{Giamarchi95};
on these we concentrate in the following.
\par
We investigate the standard model of spinless fermions on a ring,
with nearest-neighbor interaction $V$ (in units of the hopping amplitude),
the lattice constant is unity, $M$ denotes the number of sites 
(i.e. the system size),
and $N$ the fermion number.
We restrict ourselves to the case of  half-filling, $N=M/2$,
and consider on-site disorder.
The model is described by the Hamiltonian
$\HH = \HH_K + \HH_V + \HH_I$, with
\begin{equation}
 \HH_K = -\sum_{j} \left( \ee{\imath\varphi} c^{+}_{j+1}\,c^{}_{j}
         + \mbox{h.c.} \right),
\end{equation}
where $\varphi = \phi/M$,
\begin{equation} 
 \HH_V = V \sum_{j} n_j n_{j+1},
\end{equation}
 and
\begin{equation}
 \HH_I = \sum_{j} \epsilon_j n_j.
\end{equation}
\par
In contrast to the ``generic'' defects described by the random on-site 
energies $\{\epsilon_j\}$, it is also possible to construct integrable
models \cite{Schmitt95} with ``transparent'' impurities,
which decrease the phase sensitivity even though there is
no wave-function localization.
The distinction between ``integrable'' and ``non-integrable'' is also
most important for the temperature dependence of the Drude weight, 
$D$ \cite{Castelle95}.
We remark that the clean case is well studied by exact methods
(see e.g.\ \cite{Hamer87,Shastry90}).
\section*{The DMRG-Algorithm}
The Density Matrix Renormalization Group algorithm, introduced by White 
\cite{White92},
is a numerical technique which allows reliable results for one-dimensional 
quantum lattice
models, of a size (up to a few hundreds sites) much larger than
accessible by exact diagonalization methods,
to be obtained.
The algorithm can, in short, be characterized
as a ``projected diagonalization'', where the subspace to be projected onto 
is determined by the 
most probable eigenstates of a density matrix. Start, for example, with a
 reasonable representation
of a $s$-site system, using $m$ relevant states. Then add one site, 
$s \rightarrow s+1$,
and supplement the 
system by an ``environment'', namely the sites $s+2,\ldots, 2(s+1)$.
The basis of $(2m)^2$ states formed in this way is used to determine  the 
ground state of 
$\HH_{2s+2}$.
(The factor ``two'' appears here since we have two states for each site.)
The density matrix of the ``system'' determines the $m$ most important states, 
onto  
which all relevant operators are projected. 
Then add another site, $s+1\rightarrow s+2$, and proceed. 
\par
The result of this ``infinite lattice'' algorithm is the basis for the 
``finite lattice''
algorithm, which is needed in order to treat non-reflection symmetric systems.
(i) The infinite lattice algorithm is used to find a representation of all 
$s$-sites systems with
$s$ up to $M-3$.
(ii) Then consider the sites $1,\ldots,s$ as system, add two sites, and take 
as environment
the $s+3,\ldots,M$ system.
(iii) Proceed as described above. This process can be characterized
as ``sweeping through the lattice''.
With $m$ up to several hundred, we have achieved sufficient accuracy,
even for questions as subtle as the phase sensitivity.
The method works best for open boundary conditions, though there is no
general problem (except enhanced computing time) 
to include twisted boundary  conditions.
Since the DMRG is a local method, disorder is easily included,
though disorder averages of course require considerable computing time.
For example, the data given in Fig.~\ref{Fig:PSr} are based on roughly 300 CPU 
days on a
high-end workstation.
Further details, e.g. the adaption to non-reflection symmetric models, 
are discussed 
in \cite{Schmitt96}.
\par
\section*{Friedel Oscillations}
The decay of the density oscillations induced by a
defect is a long-standing problem in solid state physics.
This phenomenon, called Friedel or Ruderman-Kittel oscillations 
(depending on the context),
is closely related to the singularity in the response function 
for wave-vectors
close to $2k_F$.
It is expected that asymptotically, the induced density decays as
\begin{equation}
 \delta n(x) \sim \frac{\cos\left(2k_F x + \eta_F\right)}{x^\delta} 
\label{eq:F1}
\end{equation}
Using the DMRG, we have computed $\delta n(x)$ for a system of 200 sites 
and various
interaction strengths and, as a test for the accuracy of our calculation, 
for systems with $M=500$
and $V=\pm 1,2$.
The impurity is chosen antisymmetrically for technical reasons, 
$\epsilon_1=-\epsilon_M$.
We consider a half-filled band, i.e.\ $2k_Fx_j=\pi j$,
where $j$ is the distance (in units of the lattice spacing) from the defect.
A sample of our results is shown in Fig.~\ref{Fig:F1},
where we plot, on a logarithmic scale, the magnitude $|\delta n_j|$ versus 
distance.
Clearly, it is possible to extract the exponent $\delta$ without difficulty.
We emphasize that the algebraic decay starts already at a few lattice site 
sites.
Typically, we have used a basis of $m=120$ (200) states for the $M=200$ (500)
 system,
and performed four sweeps through the lattice.
\par
The exponent $\delta$ as a function of the nearest-neighbor interaction $V$
is given in Fig.~\ref{Fig:F2}.
The exponent $\delta$ decreases with increasing repulsive interaction,
and increases with attractive interaction, compared to the value for
non-interacting fermions, $\delta=1$ (one dimension!).
\par
Qualitatively, this trend agrees with the prediction \cite{Egger95} 
based on the Luttinger liquid.
In a recent preprint \cite{Wang96}, $\delta$
was related to the ``dressed charge'' of the (clean) model, with the result 
$\delta=Z^2=\pi/4\eta$, where $\eta$, related to $V$ through 
$V=-2 \cos{(2\eta)}$,
parameterizes the interaction.
The expression $\delta=\pi/4\eta$ is also shown in Fig.~\ref{Fig:F2},
and is in almost perfect agreement with our numerical data,
except for $V>1$ where we find that the oscillations decay
more weakly than predicted.
This seems to be related to the crossover (for a weak impurity, and $V>0$) 
found
in \cite{Egger95}, i.e.\ for the system sizes studied we may not yet be
in the asymptotic regime.
We have preliminary results showing  that for a strong
impurity, $\delta$ tends to increase towards the asymptotic
result given in \cite{Wang96}.
\section*{Phase Sensitivity}
Concerning the phase sensitivity of the ground state energy, 
let us recall the free electron result, for odd $N$:
\begin{equation}
 \left[ E(\phi) - E(0) \right]_{0} = \frac{\hbar v_F}{2\pi L} \phi^2
\end{equation} 
to be continued periodically outside the interval $-\pi\ldots\pi$.
For an even number of particles, $\phi\rightarrow\phi-\pi$ in this equation.
In our units, and for a half-filled band, we have $\hbar v_F/L\rightarrow 2/M$;
thus $\DE_0 = (-1)^N [ E(0) - E(\pi) ]_{V=0} = \pi/M$
for the clean, non-interacting system.
\par
Disorder is introduced by taking the on-site energies $\{\epsilon_j\}$ 
as random quantities,
uniformly distributed over the range $-W/2\ldots W/2$, which corresponds 
for free fermions to a conductance localization length
$\xi_0 \approx 105/W^2$, i.e.\ the average conductance decreases 
$\langle g\rangle\sim \exp{(-2M/\xi_0)}$ for a wire of length $M$ 
\cite{Kappus91}.
We considered $W=2$ only, hence $\xi_0 \approx 26$.
Naturally, we first studied non-interacting fermions, using 500 samples,
i.e.\ 500 different realizations of the disorder for each system size.
Interestingly, we find $\Eaverage{\DE_0} \sim \Eaverage{g}^{1/4}$,
but we also see that the $\DE$-distribution is rather asymmetric,
in contrast to the distribution of $\ln(M\DE)$.
The average $\Eaverage{\ln(M\DE_0)}$ decreases as $\mbox{\it const}-M/\xi$, 
with $\xi\approx 29$,
close to the conductance localization length.
\par
In Fig.~\ref{Fig:PSr} we present our results, i.e.\ $M\DE$ versus $M$,
for a repulsive interaction, $V=1.0$.
The 40 (50,60) sites systems have been calculated using $m=190$ (375,375) 
states
and performing three finite lattice sweeps.
The accuracy is better than $10^{-5}$ for the groundstate energy 
of the 60-sites systems.
The phase sensitivity, $\DE$, is positive for all samples.
The dashed line, obtained by fitting $\Eaverage{\ln(M\DE)}$ as
described above, represents $-M/\xi$ with $\xi\approx 14$, 
about half of the free-fermion value.
The error in $\xi$ may be about 20\%.
(We feel that for $M=60$, the number of samples used, $\approx 50$,
 may not be sufficient.)
The short-dashed line shows, for comparison, the decay of 
$\Eaverage{\ln(M\DE)}$
for the non-interacting case.
\par
In contrast, the phase sensitivity  is strongly enhanced for an attractive 
interaction,
as can be seen from Fig.~\ref{Fig:PSa} ($V=-1.0$).
The dashed line, again, is obtained by fitting $\Eaverage{\ln(M\DE)}$ with 
$-M/\xi$, leading to $\xi\approx 100$.
As an example, we have studied a 40-sites system in more detail,
for a repulsive interaction ($V=1.2$).
Using $\approx350$ samples, we find a reasonably smooth distribution of
$\ln(M\DE)$-values, as shown in Fig.~\ref{Fig:PSH}.
These data can be fitted well with the corresponding Gau{\ss} curve, 
computed from the average and the variance as given in the 
figure caption;
these values were computed from their definition.
\section*{Conclusions}
In conclusion, we have demonstrated that using the DMRG-algorithm,
it is possible to obtain very accurate results for the ground-state
properties of interacting one-dimensional systems with defects, as e.g.\ is
apparent from the comparison of numerical and analytical results for the decay
of the Friedel oscillations.
In particular, system sizes of a few hundred sites are sufficient to obtain
asymptotic results,
except when the decay is very slow, i.e.\ $\sim x^{-1/2}$ or slower.
Our results for the phase sensitivity are consistent with what is expected for
spinless fermions (see \cite{Giamarchi95} and reference therein).
The attractive and the repulsive ground states both contain density 
fluctuations,
but the attractive ground state also contains superconducting fluctuations.
The latter screen the disorder, leading to an increase of the phase senstivity,
i.e.\ a localization length larger than in the non-interacting case, 
for an attractive interaction.
On the other hand, the phase senstivity is reduced for a repulsive interaction.
As pointed out in \cite{Giamarchi95}, however,
these trends are just opposite to what should be expected for ``realistic'' 
models, i.e.\ models
which tend to homogenize the density for a repulsive interaction
(as e.g.\ is the case for the Hubbard model).
We plan to study this question further.
We are not aware, however, of any evidence that the phase sensitivity can 
become larger 
than in the clean, non-interacting case.
\section*{Acknowledgements}
We thank Karen Hallberg, Reinhold Egger, and Peter Schwab for stimulating 
discussions.
The calculations were performed on an IBM SP2 at the Leibniz-Rechenzentrum 
Munich.
We thank Christian Schaller and  Matthias Brehm from the LRZ for the support 
concerning
the SP2 cluster. 
Work supported by the Deutsche Forschungsgemeinschaft
(Forschergruppe HO 955/2-1).
\newpage
\np
\begin{figure}
\caption{Decay of the Friedel oscillations induced by weak impurities,
 located summetrically at the ends of the chain 
 ($\epsilon_1=-\epsilon_M=0.01$).
 The increase for large $j$ arises due to the finiteness of the chain.
 The calculations are performed at half filling, $N=M/2$, 
 keeping  $m=120$ (200) states
 per block for $M=200$ (500).
 The amplitude of the oscillation vanishes for $\epsilon_1\rightarrow0$.
 }
\label{Fig:F1}
\end{figure}
\np
\begin{figure}
 \caption{The exponent $\delta$ vs. interaction,
  for the same impurity as in Fig.~\ref{Fig:F1}.
  The continous line is the asymptotic result \protect\cite{Wang96}.
 } 
 \label{Fig:F2}
\end{figure}
\np
\begin{figure}
 \caption{ Phase sensitivity of the ground state energy vs. system size, 
 for a repulsive interaction ($V=1.0$, $W=2.0$, $N=M/2$).
 This value of $W$ corresponds to $\xi_0\approx 26$ for the non-interacting
 case. The decay length is $\xi\approx 14$.
 }
 \label{Fig:PSr}
\end{figure}
\np
\begin{figure}
 \caption{Phase sensitivity versus system size for an attractive interaction 
  ($V=-1.0$, $W=2.0$, $N=M/2$).
  The decay length is $\xi\approx100$.
 }
 \label{Fig:PSa}
\end{figure}
\np
\begin{figure}
 \caption{ Distribution of $\ln(M\DE)$ for $V=1.2$, $M=40$,
  on the basis of 353 samples.
  We find $\Eaverage{ \ln M\DE } = -2.54$ and $\sigma=1.5$ from our data.
  The continous line is the corresponding Gau{\ss} curve.
 }
 \label{Fig:PSH}
\end{figure}
\end{document}